\documentclass[twocol]{ametsocV6.1}

\title{Are Deep Learning Methods Suitable for Downscaling Global Climate Projections? An Intercomparison for Temperature and Precipitation over Spain}

\authors{Jose González-Abad,\aff{a}\correspondingauthor{Jose González-Abad, gonzabad@ifca.unican.es} 
and José Manuel Gutiérrez\aff{a}
}

\affiliation{\aff{a}{Instituto de Física de Cantabria (IFCA), CSIC-Universidad de Cantabria, Santander, Spain}
}
\abstract{Deep Learning (DL) has shown promise for downscaling global climate change projections under different approaches, including Perfect Prognosis (PP) and Regional Climate Model (RCM) emulation. Unlike emulators, PP downscaling models are trained on observational data, so it remains an open question whether they can plausibly extrapolate unseen conditions and changes in future emissions scenarios. Here we focus on this problem as the main drawback for the operationalization of these methods and present the results of an intercomparison experiment to evaluate the performance and extrapolation capability of existing models using a common experimental framework, taking into account the sensitivity of results to different training replicas. We focus on minimum and maximum temperatures and precipitation over Spain, a region with a range of climatic conditions with different influential regional processes. We conclude with a discussion of the findings, limitations of existing methods, and prospects for future development.} 

\begin{document}

\maketitle

%
%
%
\statement
Deep learning techniques have recently emerged as a promising approach to enhance the spatial resolution of coarse climate models, a process known as statistical downscaling. In this study, we review current methods and evaluate two popular models, DeepESD and U-Net, to assess their ability to project future climate changes over Spain, a region characterized by diverse and complex climates. While both models perform well with historical data, their effectiveness in projecting future periods varies. DeepESD demonstrates particular promise when tailored to focus on extreme events, though challenges persist in accurately modeling such events under changing climate conditions. This work provides guidance for advancing climate downscaling with deep learning, identifying key areas for further research.

%
%


\section{Introduction}

Global and Regional Climate Models (GCMs and RCMs, respectively) simulate the evolution of Earth's climate on different spatial and temporal scales by numerically solving the equations governing the dynamics of the relevant climate processes \citep{chen_framing_2021-1}. These models are used to produce future climate projections under different greenhouse gas emission scenarios \citep{eyring_overview_2016}. Whereas GCMs run at a global scale, RCMs operate regionally, nested to the GCM output over a limited area of interest, providing higher resolution and solving additional regional processes (this process is known as dynamical downscaling). The resulting projections are the main source of information to characterize the physical risk to climate change in assessment, impact and adaptation studies. Due to computational and physical limitations, the spatial resolution of these models is relatively coarse ($\sim100$ km and $\sim10$ km for state-of-the-art GCMs and RCMs, respectively) for regional and local applications. Approximately $40\%$ of users, however, demand climate change projections at a spatial resolution of 1 km or finer \citep{rossler_challenges_2019}. Statistical downscaling was introduced as a cost-effective methodology to bridge this gap by learning from data a downscaling function between the coarse large-scale atmospheric model outputs e.g., GCM, RCM, or reanalysis (predictors) and the local observations for the variable(s) of interest, e.g. temperature or precipitation (predictands) \citep{gutierrez_intercomparison_2019}.

Deep Learning (DL) \citep{goodfellow_deep_2016,prince_understanding_2023} has recently emerged as a promising (statistical) downscaling technique with the capacity to learn complex spatiotemporal relationships from data (for a recent review see \citet{rampal_enhancing_2024}). These models have been  used in a number of downscaling applications following different approaches, such as simple Super Resolution (SR) downscaling \citep{vandal_deepsd_2017, sha_deep_2020a,sha_deep_2020b}, bias adjustment \citep{chaudhuri_cligan_2020, francois_adjusting_2021}, a combination of both \citep{lin_deep_2023,pan_future_2024}, and more advanced Perfect Prognosis (PP) downscaling \citep{bano_downscaling_2022} and RCM emulation \citep{doury_regional_2023,van_deep_2023,bano_transferability_2024}. The latter two are the most comprehensive approaches for statistical downscaling, since the use of large-scale predictors (e.g., specific humidity and wind components at different pressure levels) allow exploiting the relationship between large-scale synoptic patterns and regional variables, thus capturing potential mechanisms for regional added value. These methods constitute an active topic of research due to their potential application to generate large downscaled ensembles from multiple model and scenario projections. In the following, for the sake of simplicity, we will refer to emulators and downscaling to indicate RCM emulators and PP statistical downscaling approaches, respectively. 

A fundamental difference between these two approaches is that, while RCM emulators learn the downscaling function in the model space (i.e. using predictors and predictands from the GCM/RCM and RCM, respectively), downscaling models learn in the observational space (using reanalysis and observational gridded or local station data). This condition confers downscaling models greater flexibility regarding the spatial resolution of predictands, since they can use any available observational dataset. However, this flexibility can also be a limitation. This arises from the potential for multiple observational datasets to exist for the same region, each offering a different approximation of the true, yet unknown, reality. These differences may depend on various aspects such as the generation technique (e.g., geostatistical techniques and reanalysis) or the resulting spatial resolution. Consequently, this may introduce an additional source of uncertainty when relying on the PP statistical downscaling approach. Furthermore, this approach requires that the downscaling function is able to plausibly extrapolate unseen conditions and changes in future emissions scenarios. This is a major drawback, since the lack of generalization can produce trend artifacts transforming the climate change signal. This problem remains an open question and hampers the operationalization of these methods, particularly in the case of the complex DL downscaling methods with a huge number of parameters lacking comprehensive explainability. 

Here we focus on this problem and present the results of an intercomparison experiment designed to evaluate the performance of  deep downscaling models and to assess their extrapolation capability. We first  conducted a literature review to identify state-of-the-art DL models used for downscaling, focusing on those applied to downscale GCM future climate projections. We found that most of the applications were based on different types of convolutional networks, fully convolutional models such as U-Net \citep{ronneberger_unet_2015}, and convolutional and dense models such as DeepESD \citep{bano_downscaling_2022}, so we used these two methods in the intercomparison. We focused on Iberia, a region with complex climatic conditions and different influential regional processes, using a gridded observational dataset for both (minimum and maximum) temperatures and precipitation with 5km spatial resolution covering the Spanish area. The intercomparison considers technical aspects such as the loss functions used in the training and the sensitivity of the results to different training replicas, thus taking into account the uncertainty coming from the use of DL techniques. 

The paper is structured as follows. In Section \ref{sec:DL4SD}, we perform an extensive literature review of papers encompassing DL models for  downscaling. In Section \ref{sec:framework}, we describe the experimental framework of this work, including the data used and the DL models trained and in Section \ref{sec:results} we present the results of the resulting experiments. Finally, in Sections \ref{sec:discussion} and \ref{sec:conclusions}, we discuss these results and conclude with the main findings of this study.

\section{Deep Learning for Statistical Downscaling}
\label{sec:DL4SD}

In the context of PP downscaling, DL models aim to learn an empirical downscaling function $f$ between the large-scale spatial predictors ($\mathbf{X}$) and the spatial predictand(s) of interest ($ \boldsymbol{y}$), using reanalysis/observational data.  These DL models are defined by a deep composition of different layers (typically convolutional and dense) of nonlinear functions involving multiple parameters (weights) which are learnt from data by minimizing a suitable loss function. Convolutional layers \citep{lecun_convolutional_1995}  capture automatically spatial patterns in the data (e.g. informative predictors and regions of influence), whereas final dense layers allow specializing the downscaling locally, introducing output-specific weights. 

A thorough revision of the literature was conducted to identify the DL methods used for  downscaling. The results are summarized in Table \ref{papersPPdownscaling}, displaying the studies in rows following a chronological order, displaying information about the region(s) and variable(s) of interest, the type of DL model, the loss function used for training and the deterministic or stochastic nature, and whether the study presents results for projections and not just training with reanalysis data. This table shows that existing studies are mostly based on convolutional models, either fully convolutional (such as U-Net) or combinations of convolutional and dense (DeepESD).

\begin{table*}
\caption{List of the most recent papers on the application of DL models for PP downscaling. The table displays the region(s) and variable(s) of interest, the type of DL model, the loss function (and the deterministic/stochastic nature, D/S) and  indicates if the study presents results for projections.}
\label{papersPPdownscaling}
\begin{center}
\scriptsize
\begin{tabular}{cccccc}
    \hline\hline
    \textbf{Ref.} & \textbf{Region} & \textbf{Var.} & \textbf{DL model} & \textbf{Loss func.} & \textbf{Projs.} \\
    \hline\hline
    \citet{misra_statistical_2018} & India and Canada & Precip. & Recurrent Neural Network & MSE (D) & No\\
    \hline
    \citet{pan_improving_2019} & United States & Precip. & \begin{tabular}[t]{@{}c@{}}Convolutional and\\dense network\end{tabular} & RMSE (D) & No\\
    \hline
    \citet{miao_improving_2019} & China & Precip. & Recurrent Neural Network & MSE (D) & No\\
    \hline
    \citet{bano_configuration_2020} & Europe & Temp. and precip. & DeepESD & NLL (S) & No\\
    \hline
    \citet{adewoyin_tru_2021} & United Kingdom &  Precip. & Recurrent U-Net & \begin{tabular}[t]{@{}c@{}} Cross-entropy\\and MSE (S)\end{tabular} & No \\
    \hline
    \citet{bano_suitability_2021} & Europe & Temp. and precip. & DeepESD & NLL (S) & Yes \\
    \hline
    \citet{sun_statistical_2021} & China & Temp. and precip. & DeepESD & NLL (S) & No \\
    \hline
    \citet{quesada_repeatable_2022} & Saxony, Germany & Precip. & U-Net & NLL (S) & No \\
    \hline
    \citet{rampal_high_2022} & New Zealand & Precip. & DeepESD & NLL (S) & No \\
    \hline
    \citet{me_multi_2022} & \begin{tabular}[t]{@{}c@{}}Southern South\\America\end{tabular} & Temp. and precip. & Dense Network & NLL (S) & Yes \\
    \hline
    \begin{tabular}[t]{@{}c@{}}\citet{hernanz_evaluation_2022a}\\\citet{hernanz_evaluation_2022b}\\\citet{hernanz_evaluation_2022c}\end{tabular} & Spain & Temp. and precip. & Dense Network & MSE (D) & Yes \\
    \hline    
    \citet{balmaceda_assessing_2022} & \begin{tabular}[t]{@{}c@{}}Southern South\\America\end{tabular} & Temp. & Dense Network & MSE (D) & No \\
    \hline
    \citet{vaughan_convolutional_2022} & Europe & Temp. & \begin{tabular}[t]{@{}c@{}}Convolutional conditional\\neural processes\end{tabular} & NLL (S) & No \\
    \hline
    \citet{olmo_statistical_2022} & \begin{tabular}[t]{@{}c@{}}Southern South\\America\end{tabular} & Precip. & Dense Network & NLL (S) & No \\
    \hline
    \citet{bano_downscaling_2022} & Europe & Temp. and precip. & DeepESD & NLL (S) & Yes \\
    \hline
    \citet{quesada_downscaling_2023} & Saxony, Germany & \begin{tabular}[t]{@{}c@{}}Temp., precip.\\and others\end{tabular} & U-Net & NLL (S) & Yes \\
    \hline
    \citet{soares_high_2023} & Iberia & Temp. and precip. & DeepESD & NLL (S) & Yes \\
    \hline
    \citet{kheir_improved_2023} & Egypt & Temp. & DeepESD & NLL (S) & Yes \\
    \hline
    \citet{gonzalez_using_2023} & North America & Temp.  & DeepESD and U-Net & MSE (D) & Yes \\
    \hline
    \citet{balmaceda_use_2024} & \begin{tabular}[t]{@{}c@{}}Southern South\\America\end{tabular} & Temp. & DeepESD & NLL (S) & Yes \\
    \hline
    \citet{bailie_quantile_2024} & \begin{tabular}[t]{@{}c@{}}New Zealand\end{tabular} & Precip. & \begin{tabular}[t]{@{}c@{}}Convolutional, dense and\\attention-based\end{tabular} & NLL (S) & No \\
    \hline
    \citet{balmaceda_regional_2024} & \begin{tabular}[t]{@{}c@{}}Southern South\\America\end{tabular} & Temp. & Dense Network & MSE (D) & Yes \\
    \hline
    \citet{hosseini_improving_2024} & Tabriz city & Temp. and precip. & \begin{tabular}[t]{@{}c@{}}Convolutional and\\dense network\end{tabular} & Not specified & Yes \\
    \hline
\end{tabular}
\end{center}
\end{table*}

The first applications focused on daily precipitation using a model composed of a set of convolutional layers and two final dense layers \citep{pan_improving_2019}, and recurrent variations taking into account the temporal dimension \citep{misra_statistical_2018,miao_improving_2019}. These models were trained to minimize the Mean Square Error (MSE) loss function (or its squared root, RMSE), which quantifies the error as the average squared difference between the prediction $\boldsymbol{\hat{y}}$ generated by the DL model and the true value $\boldsymbol{y}$ 

\begin{equation}
    J(\boldsymbol{\theta}) = \frac{1}{N} \sum_{i=1}^{N} \left(\boldsymbol{\hat{y}}_i-\boldsymbol{y}_i\right)^2,
\end{equation}

\noindent where N corresponds to the number of samples composing the dataset. Note that, in the context of downscaling, loss functions are generally computed independently for each grid point in the predictand, thereby measuring the error in a pixel-wise manner

\citet{bano_configuration_2020} introduced the DeepESD model for downscaling temperature and precipitation over Europe, consisting of three convolutional layers followed by a (linear) dense layer. This model has been applied in several following studies, such as in \citet{bano_suitability_2021,bano_downscaling_2022}, where results applying these methods to produce downscaled GCM future projections are presented for the first time, reporting that they are suitable for the generation of plausible projections in future scenarios. One of the main characteristics introduced by the DeepESD model is the loss function used in the learning process. The MSE loss function causes the model to underrepresent extremes, which can be problematic for variables such as precipitation, highly characterized by these events (e.g., heavy rainfalls). To account for this, following previous studies \citep{dunn_occurrence_2004,cannon_probabilistic_2008}, DeepESD explicitly models the conditional probability distribution using an stochastic loss function  minimizing the Negative Log-Likelihood (NLL) of the target distribution. For instance, for temperature, DeepESD predicts the parameters $\mu$ and $\sigma$ of a Gaussian distribution for each grid point in the predictand and for precipitation the parameters $p,\alpha$ and $\beta$ of a combination of Bernoulli and gamma distributions for occurrence and amount, respectively, as follows:

\begin{equation}
    P(y|\mathbf{X};p,\alpha,\beta) = 
    \begin{cases}
      1-p, & \text{if}\ y < 1 \\
      \frac{p}{\Gamma(\alpha)\beta^\alpha}y^{\alpha-1}e^{-y/\beta}, & \text{if}\ y \geq 1 ,
    \end{cases}
\end{equation}

\noindent where $p$ corresponds to the probability of rain, $\alpha$ and $\beta$ to the shape and scale parameters of a gamma distribution and $\Gamma$ to the gamma function. This allows sampling from the estimated distribution and, thus, representing extremes. Note that in the case of temperature, minimizing the MSE loss function is equivalent to maximizing the likelihood of a Gaussian distribution with a fixed $\sigma$ \citep{goodfellow_deep_2016}.

DeepESD was rapidly adopted in various applications  across different spatial domains such as China \citep{sun_statistical_2021}, New Zealand \citep{rampal_high_2022}, Egypt \citep{kheir_improved_2023}, southern South America \citep{balmaceda_use_2024} and Iberia \citep{soares_high_2023}. These studies show that the DeepESD model outperforms standard statistical downscaling techniques, although in some cases it can amplify the climate change signal of the GCM \citep{balmaceda_assessing_2022}. DeepESD has also inspired other works; for instance, \citet{hosseini_improving_2024}  intercompare similar DL models for the downscaling on Tabriz city (Iran) with successful results when projecting various GCMs. Similarly, \citet{bailie_quantile_2024}  propose an ensemble of DeepESD-like models, each of these trained it over a different partition of the precipitation distribution, allowing them to specialize and better capture extremes.

In parallel to the development of DeepESD, some studies adopted the fully convolutional U-Net architecture \citep{ronneberger_unet_2015}, which was originally designed for image segmentation tasks. It is composed of two different blocks, encoder and decoder, with convolutions and transposed convolutions, respectively, and includes skip connections between layers. \citet{quesada_repeatable_2022} applied this model in Germany using the stochastic version, minimizing the NLL of Gaussian and Bernoulli-gamma distributions for temperature and precipitation, respectively. The performance of these U-Nets was compared to DeepESD, with the former exhibiting slightly better results in the specific studied region. U-Net models have also been extended to incorporate the temporal dimension of large-scale data for precipitation downscaling over the United Kingdom \citep{adewoyin_tru_2021}. Despite achieving satisfactory performance, this model tends to underestimate extreme precipitation values, a recurring challenge in DL-based downscaling models. A significant factor contributing to this issue in this specific work may be the choice of the MSE as the loss function for modeling the amount. Applications of more advanced DL models have involved convolutional conditional neural processes capable of downscaling to coordinates not seen during the training phase \citep{vaughan_convolutional_2022}. As previous works, this model is constructed following the NLL loss function developed for DeepESD models.

Despite the progress made with CNNs, recent works still involve the intercomparison of simpler dense networks. For example, \citet{hernanz_evaluation_2022a,hernanz_evaluation_2022b,hernanz_evaluation_2022c}, compare dense networks for temperature and precipitation downscaling over Spain. In terms of future projections, the authors evaluate these models in a pseudo-reality experiment and find that neural networks outperform other models for temperature, such as analog methods, multiple linear regression and support vector machines, while all machine learning models achieve similar results for precipitation. Additional analyses have been carried out for the southeastern South America region \citep{me_multi_2022,olmo_statistical_2022,balmaceda_regional_2024}. 

Finally, \citet{gonzalez_using_2023} compared the DeepESD and U-Net models for downscaling temperature in North America using eXplainable Artificial Intelligence (XAI) techniques. They found that DeepESD can learn spurious pattern when trained on a large region with very contrasting climates. 

\section{Experimental Framework, Models and Data}
\label{sec:framework}

\subsection{Area of Study}
In this study we focus on the peninsular Spain ($36^\circ$N-$44^\circ$N and $9.5^\circ$W-$3.5^\circ$E), the Spanish territory located within the Iberian peninsula. This region is situated in the Mediterranean basin, a region heavily
affected by climate change, with rising temperatures, altered precipitation patterns, and
increased frequency of extreme weather events \citep{hoerling_increased_2012,russo_synergy_2019,cos_mediterranean_2022}. In addition, this region exhibits diverse climatologies and a complex orography. This hampers the regional assessment of climate change, key for understanding the diverse impacts of climate change.  All these characteristics make this region interesting for the evaluation of downscaling techniques.

\subsection{Observational Data: ROCIO-IBEB}
As predictand (i.e., the observational data used as the target for model training), we choose the daily minimum and maximum temperature and accumulated precipitation (see Table \ref{tab:list_predictor_predictand} for further details) from the observational dataset ROCIO-IBEB 5km provided by the Agencia Estatal de Meteorología (AEMET) \citep{peral_serie_2017}. This dataset provides high resolution data ($0.05^\circ$ horizontal resolution) for Spain and incorporates over 2000 ground stations which are assimilated using a 3D variational scheme imposing physical consistency. This large number of ground stations is crucial for a geostatistical dataset of this nature. Furthermore, the use of this dataset as a foundation for the latest National Plan for Climate Change Adaptation (PNACC), Spain's strategic framework for addressing climate change challenges, supports its reliability and quality.

\begin{table*}
\caption{Surface variables of interest (predictands) and large-scale variables used as predictors for the downscaling methods.}
\label{tab:list_predictor_predictand}
\centering
    \begin{tabular}{l|lll}
    \hline
                 & \textbf{Variable} & \textbf{Level} & \textbf{Unit} \\ \hline
                 & Minimum temperature & Surface & $^{\circ}$C \\
     Predictands & Maximum temperature & Surface & $^{\circ}$C \\
                 & Accumulated precipitation & Surface & mm/day \\               
   \hline
                 & Air temperature & 850, 700 and 500 hPa & $^{\circ}$C (K) \\
                 & Specific humidity & 850, 700 and 500 hPa & kg kg$^{-1}$ \\
      Predictors & Meridional wind component & 850, 700 and 500 hPa & m s$^{-1}$ \\
                 & Zonal wind component & 850, 700 and 500 hPa & m s$^{-1}$ \\
                 & Mean sea level pressure & - & Pa \\  \hline
\end{tabular}
\end{table*}

\subsection{Model Data: Reanalysis Predictors}
Following the PP approach, we represent the state of the atmosphere by selecting as predictors the set of daily large-scale variables listed in Table \ref{tab:list_predictor_predictand}. This selection is inspired by established work in downscaling literature \citep{huth_statistical_2002,huth_downscaling_2005,gutierrez_reassessing_2013} and the recommendations from VALUE \citep{maraun_value_2015}, a detailed evaluation framework for statistical downscaling techniques in the context of climate change. These variables are obtained from the ERA5 reanalysis \citep{hersbach_era5_2020} developed by the European Centre for Medium-Range Weather Forecasts (ECMWF), with a  $0.25^{\circ}$ horizontal resolution. The predictors were re-gridded using conservative interpolation to a $1.5^{\circ}$ resolution, to better match the coarse resolution of the GCMs. The predictors span a wider spatial domain ($23.5^\circ$N-$68.5^\circ$N and $39^\circ$W-$22.5^\circ$E) with the objective of properly capturing large-scale phenomena influencing the downscaled surface variables (see Figure \ref{fig:dl_models} for an illustration of the predictor domain). Finally, before passing these predictor variables to the DL model we standardize them grid point by grid point to have mean 0 and standard deviation 1, thus avoiding discrepancies in the scale of different variables and accelerating convergence \citep{goodfellow_deep_2016}. 

\subsection{Model Data: GCM Historical and Future Projections}
\label{sec:framework:gcm}

To generate regional projections in future periods, we use the GCMs listed in Table \ref{tab:GCMlist}. These models are part of the latest iteration of the Coupled Model Intercomparison Project (CMIP6), an initiative aimed to evaluate and compare multi-model ensembles composed of different GCMs generated from different Shared Socioeconomic Pathway (SSP) scenarios \citep{riahi_locked_2015, chen_framing_2021-1}. In addition, the selected GCMs are among those recommended by EURO-CORDEX for downscaling from CMIP6 \citep{sobolowski_euro_2023}. We used information from both historical and future SSP3.70 scenario, which represents a high emission scenario and allows to assess extrapolation capabilities to unseen climatic conditions. We selected this scenario because, first, it is recommended by the EURO-CORDEX initiative to represent a high-emissions pathway \citep{sobolowski_euro_2023}, and second, it best reflects the outcome projected under current climate policies and conditions \citep{hausfather_emissions_2020}.

\begin{table*}
\caption{Global Climate Models (GCMs) downscaled in this work, including their references, modeling centers, horizontal resolutions and the variables downscaled with each of them.} 
\label{tab:GCMlist}
\centering
    \begin{tabular}{cccc}
        \hline
            \textbf{Name} &  \textbf{Institution} & \textbf{Resolution} & \textbf{Downsc. vars.}\\
        \hline
            \begin{tabular}[t]{@{}c@{}}EC-Earth3-Veg \\ \citep{doscher_ec-earth3_2022}\end{tabular} & EC-Earth Consortium & $\sim 80$ km & \begin{tabular}[t]{@{}c@{}}Minimum and\\maximum temp \\and precip.\end{tabular} \\
        \hline
            \begin{tabular}[t]{@{}c@{}}MPI-ESM1-2-LR \\ \citep{muller_higher_2018} \end{tabular} & \begin{tabular}[t]{@{}c@{}} Max Planck Institute for\\Meteorology (Germany) \end{tabular} & $\sim 100$ km & \begin{tabular}[t]{@{}c@{}}Minimum and\\maximum temp.\end{tabular} \\
        \hline
            \begin{tabular}[t]{@{}c@{}}CMCC-CM2-SR5 \\ \citep{cherchi_global_2019} \end{tabular}  & \begin{tabular}[t]{@{}c@{}}Centro euro-Mediterraneo sui\\Cambiamenti Climatici (Italy)\end{tabular} & $\sim 100$ km & Precip. \\
        \hline
\end{tabular}
\end{table*}

Taking into account the PP assumptions, and following previous works \citep{bano_downscaling_2022,risser_bias_2024}, we perform a bias adjustment of the GCM predictors to increase the distributional similarity with their counterpart ERA5 reanalysis predictor fields. In particular, a signal-preserving adjustment of the monthly mean and variance of the GCM predictors $x$ working on a calendar month basis is computed as follows:

\begin{equation}
    x_{f}^{\prime \, m} = \frac{x_{f}^{m} - \Delta^m - \mu_h^m}{\sigma_h^m} \sigma_e^m +\mu_e^m + \Delta^m, 
\label{eq:xai:biasAdjustent}
\end{equation}

\noindent where $\Delta^{m} = \mu_f^{m} - \mu_h^{m}$ corresponds to the climate change signal for month $m$ of the GCM, computed by subtracting the future ($f$) and historical ($h$) monthly means computed over the years corresponding to each period, whereas $\mu_e^m$ and $\sigma_e^m$ correspond to the mean and standard deviation of the reanalysis dataset, respectively. Consequently, this adjustment is calibrated using GCM historical data and the corresponding ERA5, and then applied to the future period. As can be seen, to preserve the climate change signal, this signal ($\Delta^{m}$) is extracted from the future period (mean difference between the future and the historical period) and added again after adjusting the daily data, working also on a calendar month and period basis. Finally, the bias-adjusted GCM predictors are standardized before passing them to the DL model in the same way as ERA5 predictors.

\section{Deep Learning Models}
\label{sec:DL_models}

\begin{figure*}
    \noindent\includegraphics[width=0.95\linewidth]{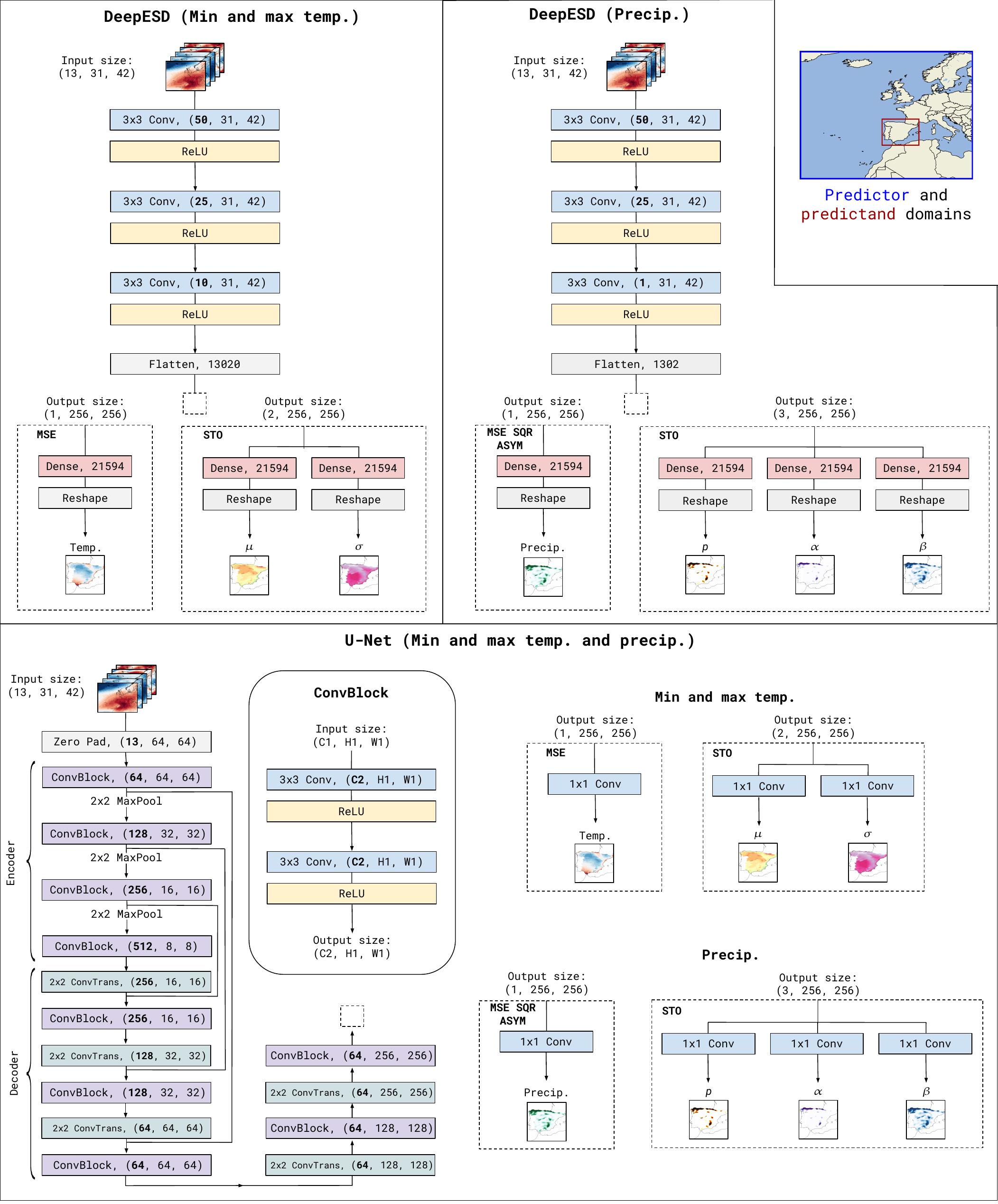}
    \caption{Schematic view of the DeepESD (top) and U-Net (bottom) models. Due to differences in the architecture, for DeepESD we show two different versions for the two types of variables being downscaled: minimum or maximum temperature (top-left) and precipitation (top-right). For each of these models, the final set of layers specific to the loss function (and variable for the U-Net) is shown within dashed boxes. The predictor and predictand domains are also shown in the top-right part of the figure.}
    \label{fig:dl_models}
\end{figure*}

Based on the literature review in Section \ref{sec:DL4SD}, the most-common DL architectures for  downscaling were the DeepESD and U-Net models. 
Figure \ref{fig:dl_models} provides schematic views of both architectures. For DeepESD, we show two different architectures for the two types of variables intercompared: minimum or maximum temperature (top-left) and precipitation (top-right). This distinction follows \citep{bano_configuration_2020}, where the last convolutional layer for precipitation consists of one channel/kernel, in contrast to the model for temperature, which consists of 10. However, for the U-Net (bottom), there is a common part for all variables. For all these architectures, we show the final set of layers specific to each loss function (and variable for the U-Net) within a dashed box. Within each of the boxes representing the layers composing the DeepESD and U-Net architectures, we show the output size. The final output size of the model for each specific final set of layers is also shown at the top of the respective dashed box. In the following, we describe the two architectures in more detail.

\subsection{DeepESD}
The DeepESD model is composed of a mixture of convolutional and dense layers. More specifically, it is integrated by three convolutional layers with 50, 25 and 10 (1) kernels for temperature (precipitation) with Rectified Linear Unit (ReLU) activation functions \citep{glorot_understanding_2010} for the hidden layers (we refer to Figure \ref{fig:dl_models} for further details). The output of the last convolutional layer is flattened and passed as input to a final dense layer. This final layer has 21594 output neurons, which corresponds to the number of grid points to downscale. Finally, this vector is reshaped to match the size of the predictand ($256\times256$). The number of final dense layers depends on the function being minimized. For the models minimizing deterministic loss functions (MSE, SQR and ASYM losses) only one final dense layer is required as they directly output the variable to downscale. However, when minimizing the NLL (referred as STO losses), the number of dense layers is doubled (tripled) for temperature (precipitation) because we need to predict the set of parameters of the distributions estimated for each grid point in the predictand ($\mu$ and $\sigma$ for temperature, and $p$, $\alpha$, and $\beta$ for precipitation). This increase in the size of the output significantly impacts the number of parameters in the different DeepESD models, as dense layers possess the most parameters due to their densely connected nature.

\subsection{U-Net}
U-Net is a fully convolutional model composed exclusively of convolutional layers, with no dense elements. First, before passing the input to the encoder, we pad it with zeros until a spatial dimension of $64\times64$ is achieved. This is because, in the image domain, the encoder (decoder) typically reduces (increases) the spatial dimension of the data in powers of two. Therefore to maintain alignment with its original conception, we follow the same structure. Then, the encoder applies a series of convolutional layers and ReLU activation functions (ConvBlock), followed by max pooling to reduce the spatial dimension. The decoder then reconstructs the spatial dimension using transposed convolutional layers while taking advantage of the hidden features computed in the encoder through skip connections. Since the U-Net model was originally developed for image segmentation tasks where the input and output sizes are the same, we perform two additional transposed convolutions to achieve the desired $256\times256$ output size. Finally, similar to DeepESD, we apply one or multiple convolutional layers with a $1\times1$ kernel, depending on the loss function. Note that this kernel size is commonly employed at the last layer in both the image \citep{ronneberger_unet_2015} and climate domains \citep{doury_regional_2023,van_deep_2023,doury_suitability_2024}. Despite being composed of a higher number of layers (thus being deeper), the U-Net has fewer parameters than the DeepESD. This is due to the shared nature of convolutional layers, which is one of their key strengths.

\section{Model Training and Loss Functions}
\label{sec:DL_train}

Besides the architectures presented in the previous section, we also assess the importance of the loss function employed in training these models. For downscaling minimum and maximum temperatures, we focus on the MSE and stochastic (STO) loss functions. The former involves minimizing the MSE, leading the model to learn the mean conditioned on the large-scale predictors. The latter corresponds to minimizing the NLL of a Gaussian distribution, which allows us to sample from the modeled conditional distribution, rather than obtaining a single value (refer to Section \ref{sec:DL4SD} for more details on these loss functions).

For the downscaling of precipitation, we also assess the MSE as loss function. However, this loss function could lead to under-representation of extremes and lack of variability. Therefore, for this variable, in addition to the MSE, we explore a broader set of loss functions. First, as previously mentioned in Section \ref{sec:DL4SD}, we minimize the NLL of Bernoulli and gamma distributions—an approach we refer to as STO, following the naming convention used for temperature downscaling. Another strategy to address the long tails of precipitation distribution involves applying different transformations to the variable. For instance, taking the square root or logarithm of precipitation values can produce a more Gaussian-like distribution, making loss functions like MSE more suitable \citep{adewoyin_tru_2021,harris_generative_2022}. To represent this approach, we evaluate a model trained to minimize the MSE on the square root of the predictand, which we refer to as SQR. Other approaches to cope with precipitation assign higher weights to tail events when computing the loss function, thereby favoring the model to deviate from the mean and better reproduce extreme values \citep{price_increasing_2022,doury_suitability_2024}. To represent this approach we rely on the asymmetric (ASYM) loss function proposed in \citet{doury_suitability_2024}:

\begin{equation}
L_{\theta} = \frac{1}{N} \sum_{i=1}^{N} |y_i - \hat{y}_i| + \gamma^{2} \times \text{max}(0, y_i-\hat{y}_i),
\label{eq:douryOriginal}
\end{equation}

\noindent where $\gamma = G(y)$, with $G$ being the cumulative distribution of a gamma fitted to the time series of the grid point in the historical dataset. This function weights the mean absolute error by an amount proportional to how extreme is that specific precipitation value (taking into account the specific grid point) and the error itself. Notice how the weight is only applied if the model underestimates the true precipitation value. 

All DL models are trained following the same procedure, using ERA5 and ROCIO-IBEB as the predictor and predictand, respectively. The corresponding loss function is minimized using the Adam optimizer \citep{kingma_adam_2014} with a learning rate of $10^{-4}$ and a batch size of 64. We follow previous literature \citep{bano_configuration_2020} and split the observational data into a training and a test set, spanning the periods 1980-2010 and 2011-2020, respectively. To prevent the models from overfitting, we follow an early stopping strategy using a random $10\%$ split of the training data as validation data. If the loss function does not exhibit a decrease within a span of 60 epochs during this validation split of the training data, the training process is terminated. The model at the epoch with the lowest loss function value on the validation set is then selected as the final model. For each trained DL model, the training process is repeated seven times using different random seeds to initialize the model parameters, resulting in seven distinct replicas. By selecting one prediction at random from these replicas and reporting a measure of variability across them, this approach helps assess the robustness of the results with respect to the random initialization of the model parameters.

As a summary, for temperature we train $2\times2\times7$ different models corresponding to the DL architecture (DeepESD or U-Net), the loss function (MSE and STO) and the different training replicas. For precipitation, we train $2\times4\times7$ different models corresponding to the DL architecture (DeepESD or U-Net), the loss function (MSE, SQR, ASYM and STO) and the different training replicas. Note that for the STO models, the final downscaled values correspond to a random sample of the learnt conditional distributions. To evaluate the predictions from the different models, we use the metrics presented in Table \ref{tab:metrics}.

\begin{table*}
\caption{List of metrics used to evaluate the intercompared models. For each metric, the table provides a brief description, the units (with $-$ indicating no units), the variable to which it applies, and the desired value. For precipitation-related metrics, a day is considered wet if precipitation $\geq 1$ mm/day.}
\label{tab:metrics}
\centering
    \begin{tabular}{ccccc}
    \hline
    \textbf{Metric} & \textbf{Description} & \textbf{Units} & \textbf{Variable} & \textbf{Target score}\\
    \hline
    P02 bias & Bias for the 2nd percentile & $^\circ$C & Min./max. temp. & 0 \\
    \hline
    Mean bias & \begin{tabular}[t]{@{}c@{}}Bias for the mean\\Relative bias for the mean\end{tabular} & \begin{tabular}[t]{@{}c@{}}$^\circ$C\\$\%$\end{tabular} & \begin{tabular}[t]{@{}c@{}}Min./max. temp. \\Precip.\end{tabular}& 0 \\
    \hline
    P98 bias & Bias for the 98th percentile & $^\circ$C & Min./max. temp. & 0 \\
    \hline
    P99 bias & Relative bias for the 99th percentile & $\%$ & Precip. & 0 \\
    \hline
    R01 bias & Relative bias for the ratio of wet days & $\%$ & Precip. & 0 \\
    \hline
    Rx1day bias & \begin{tabular}[t]{@{}c@{}}Relative bias for the climatology of\\max. daily precipitation\end{tabular} & $\%$ & Precip. & 0 \\
    \hline
    SDII bias & Relative bias for the Simple Daily Intensity Index & $\%$ & Precip. & 0 \\
    \hline
    TNn & \begin{tabular}[t]{@{}c@{}}Bias for the annual minimum of\\daily minimum temperatures\end{tabular} & $^\circ$C & Min. temp. & 0 \\
    \hline
    TXx & \begin{tabular}[t]{@{}c@{}}Bias for the annual maximum of\\daily maximum temperatures\end{tabular} & $^\circ$C & Max. temp. & 0 \\
    \hline
    RMSE & Root mean square error & \begin{tabular}[t]{@{}c@{}}$^\circ$C\\mm/day\end{tabular} & \begin{tabular}[t]{@{}c@{}}Min./max. temp. \\Precip.\end{tabular} & 0 \\
    \hline
    Std. ratio & Ratio of std. deviations & $-$ & Min./max. temp. & 1 \\
    \hline
    Interannual var. (ratio) & Ratio of interannual variability & $-$ & Precip. & 1 \\
    \hline
\end{tabular}
\end{table*}

\section{Results}
\label{sec:results}

In this section, we intercompare the results of the DL methods trained as described in Section \ref{sec:DL_train}, for downscaling minimum/maximum temperature and precipitation over peninsular Spain. We first evaluate these models in the observational space using ERA5 as predictor and ROCIO-IBEB as predictand over the test set period (2011-2020). We then assess the performance and plausibility of the same models (trained on 1980–2010 data) when downscaling various CMIP6 GCMs under the SSP3-7.0 scenario.

\subsection{Evaluation of Model Performance}
\label{sec:results:obs}

\subsubsection{Minimum and Maximum Temperature}
\label{sec:results:obs:tas}

\begin{figure*} 
    \centering
    \noindent\includegraphics[width=\linewidth]{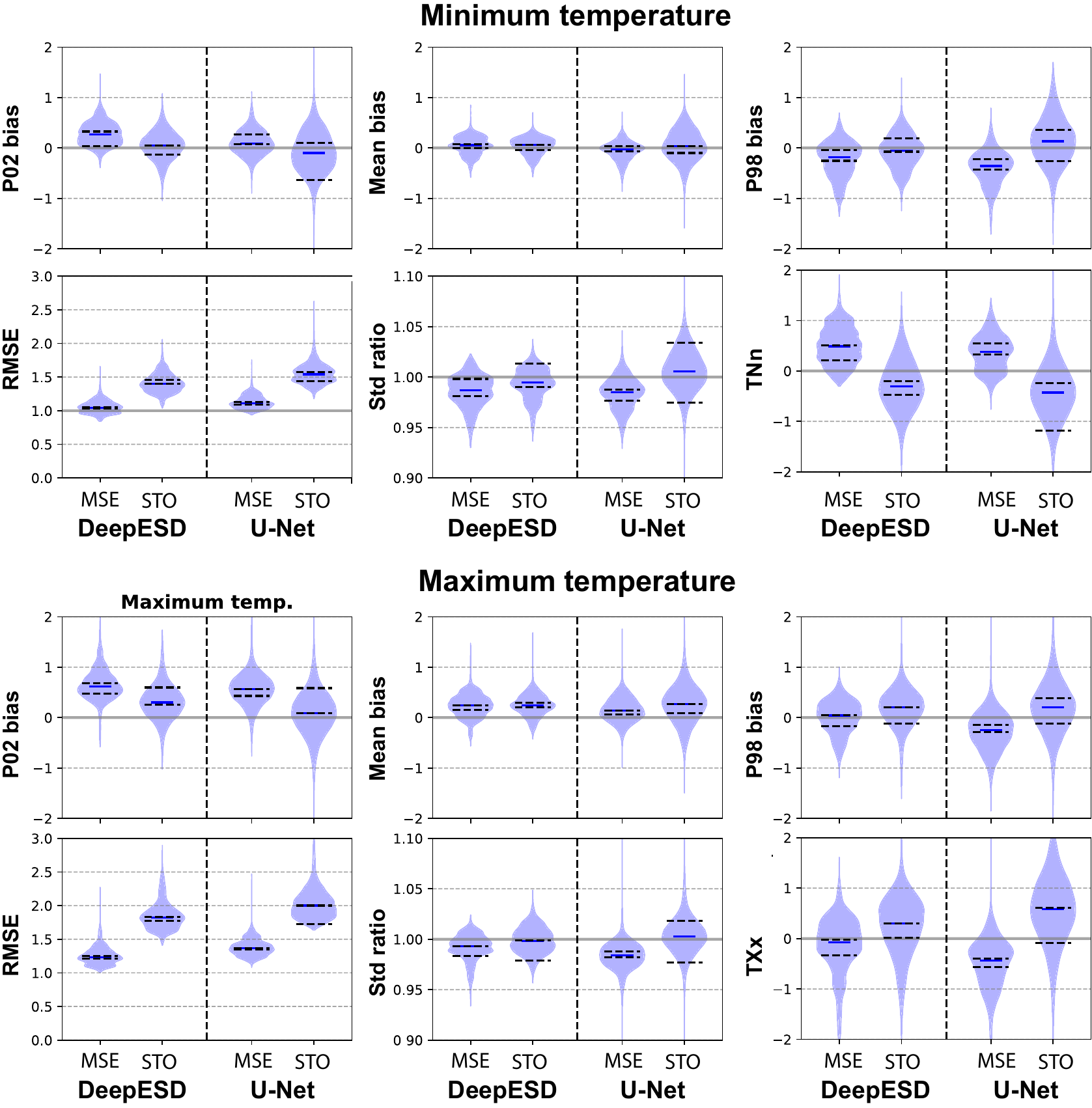}
    \caption{Evaluation results of the DL downscaling methods in the test period (2011-2020) for minimum (top) and maximum (bottom) daily temperatures using different validation metrics. Each panel includes the results for the MSE and STO versions of the DeepESD and U-Net architectures. Violins indicate the distribution of results for the different grid points for a specific training replica, with the spatial median of the corresponding replica indicated with a blue line. Black dashed lines indicate the range of variability of the corresponding spatial medians (minimum and maximum values) for the seven independent training replicas of the DL models. }
    \label{fig:fig_violin_tas}
\end{figure*}

Figure \ref{fig:fig_violin_tas} displays violin plots of the test set results for all grid points in the predictand domain. These plots show evaluation metrics for minimum (top) and maximum (bottom) downscaled temperatures using the DeepESD and U-Net models trained with MSE and STO loss functions (x-axis of each subplot). For both variables, we compute the bias of the 2nd percentile (P02 bias), the mean (Mean bias), and the 98th percentile (P98 bias), as well as the Root Mean Square Error (RMSE) and the ratio of standard deviations (Std ratio). Additionally, we compute the bias of the annual minimum of daily minimum temperatures (TNn) and the annual maximum of daily maximum temperatures (TXx) to assess the performance of the DL models regarding extreme values.

For minimum temperature, MSE-based models slightly overestimate P02 (and the minimum annual temperatures TNn, to a larger extent) and underestimate P98, not capturing the full range of variability (standard deviation ratios smaller than 1). On the other hand, the biases of STO-based are centered around zero (with the exception of  TNn, which is underestimated) and capture the observed variability, but at the cost of exhibiting larger RMSE, due to sampling from the downscaled distribution (note that by using the mean of the distribution the results would be similar to the MSE-based training). In general DeepESD exhibits better accuracy (smaller RMSE) than U-Net, but the results are overall comparable between both methods and all methods reproduce mean values. Moreover, the results from different trainings (represented in the figure showing the range of variability with black horizontal lines) indicate that the results are  robust, and do not depend on the particular training instance; the larger differences are found for the STO version of the U-Net model, where training seems to be more unstable. For the maximum temperature, the results are similar, but there is a tendency to overestimate low (P02) and mean values of the distribution.  Given their stochastic nature, STO-based models are expected to achieve better results in reproducing the extremes of the distribution. However, for minimum and maximum temperatures they exhibit a similar performance to the deterministic counterpart, with larger variability as a function of the particular training instance in the case of U-Net. 

\begin{figure*}
    \centering
    \noindent\includegraphics[width=\linewidth]{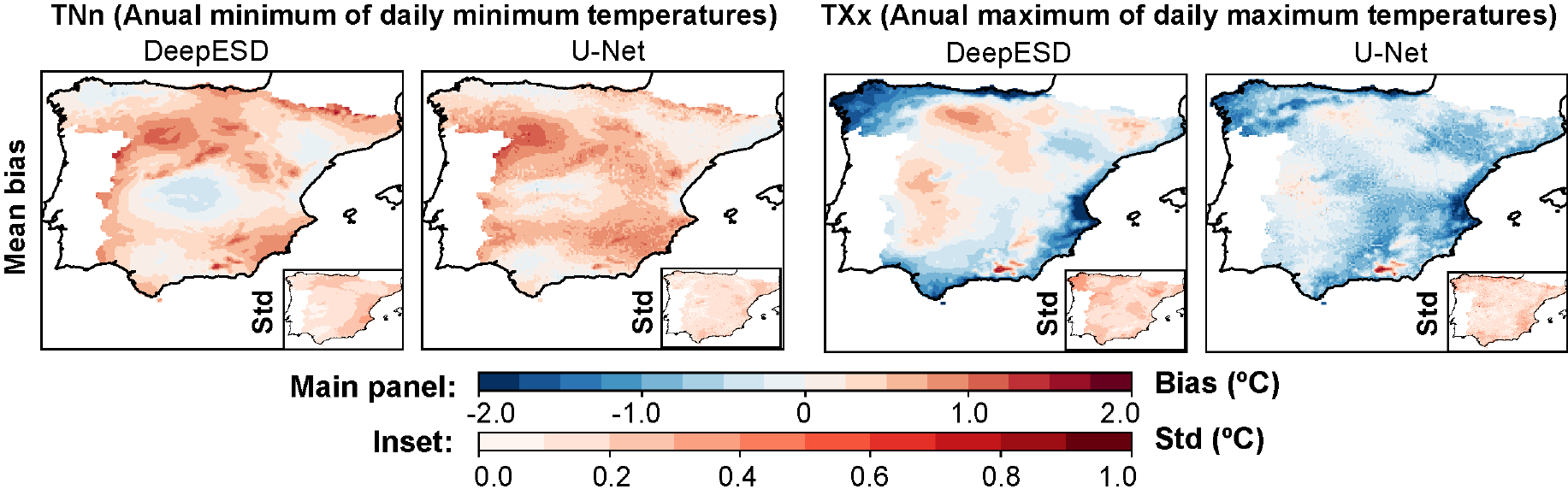}
    \caption{Mean (main panel) and standard deviation (inset) of the biases (ºC) of the seven replicas of MSE-based DeepESD and U-Net models for annual minimum of daily minimum temperatures (TNn, left) and annual maximum of daily maximum temperatures (TXx, right). }
    \label{fig:fig_extremes_tas}
\end{figure*}

Overall, the MSE-based downscaling model stands as a good performing and convenient method  to downscale minimum and maximum temperatures, so we select this model for the second part of the paper, to assess the downscaling of global climate projections. To  illustrate the spatial distribution of errors, Figure \ref{fig:fig_extremes_tas} presents, for the whole predictand domain, the TNn and TXx for the minimum and maximum temperatures for the DeepESD and U-Net models trained with the MSE loss function. These maps depict the mean bias and the standard deviation across all training replicas. For both DL models, the spatial distribution of the TNn is similar, with a large positive bias over the northwestern area of Spain. For the TXx, although several similarities can be found between the spatial patterns of the bias, DeepESD appears to overestimate TXx in the central region of Spain compared to the U-Net. Besides this, both models overestimate both TNn and TXx, for the mountain region of Sierra Nevada in the southeastern area of Spain, indicating a limitation of DL models in accurately reproducing temperatures in such regions. The standard deviation of these metrics indicates that the biases are consistent across different random initialization of the DL models, demonstrating robustness.

\subsubsection{Precipitation}
\label{sec:results:obs:pr}

\begin{figure*}
    \centering
    \noindent\includegraphics[width=\linewidth]{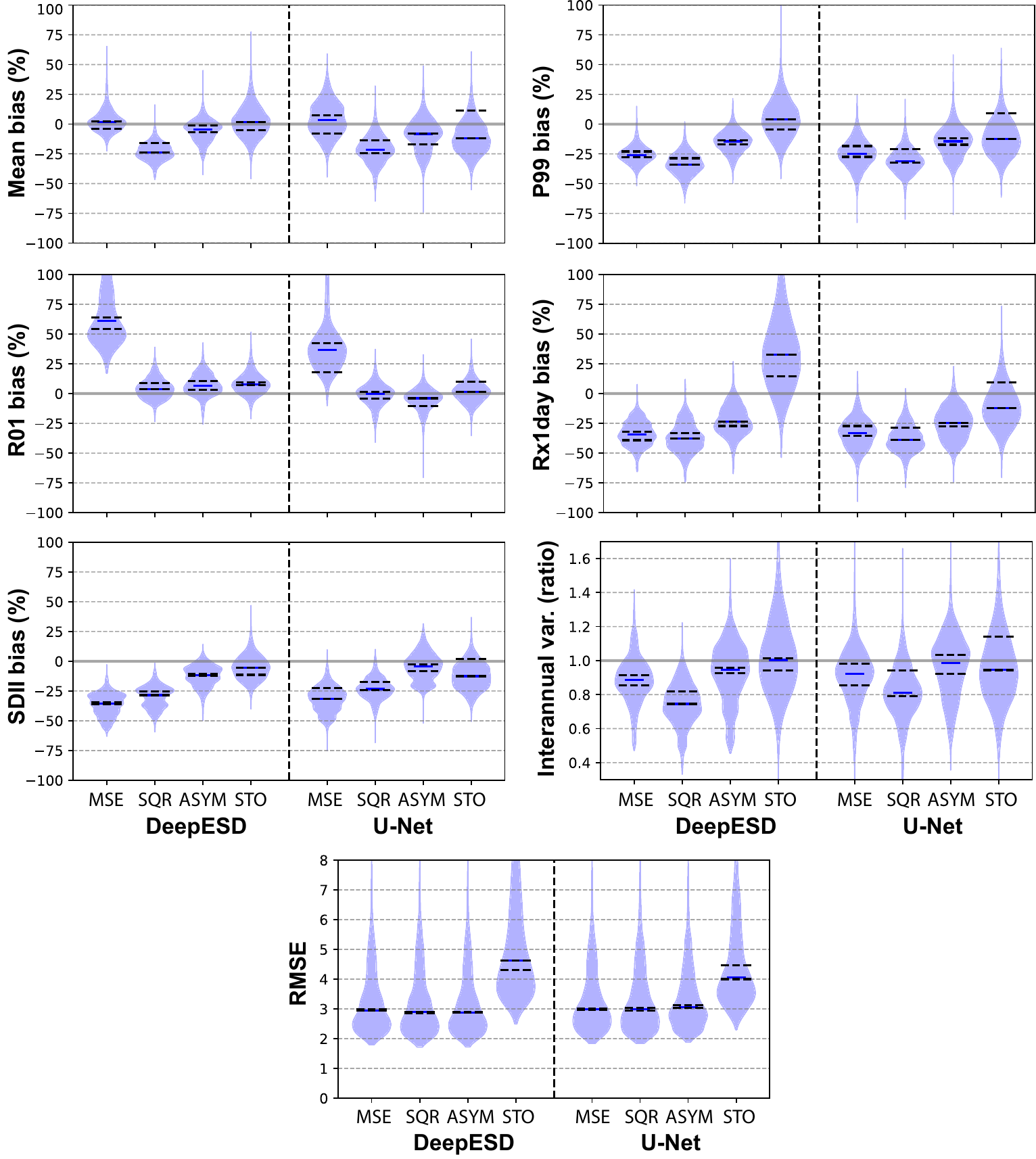}
    \caption{As Figure \protect\ref{fig:fig_violin_tas} but for precipitation. Notice that for this variable, for each DL architecture (DeepESD or U-Net), we intercompare four different loss functions: MSE, SQR, ASYM and STO.}
    \label{fig:fig_violin_pr}
\end{figure*}

Figure \ref{fig:fig_violin_pr} shows the set of evaluation metrics for precipitation. Specifically, we present violin plots for the relative bias (in \%) of the mean, the 99th percentile, the R01 (ratio of wet days, defined as days with precipitation greater than 1 mm), the Rx1day (annual maximum of daily precipitation), and the Simple Daily Intensity Index (SDII), which corresponds to the mean precipitation of wet days. In addition, we also show the ratio of interannual variability and the RMSE. These metrics are computed for both DL models for the four loss functions intercompared: MSE, SQR, ASYM and STO.

All  deterministic models exhibit a similar RMSE accuracy, smaller than the stochastic version, since the stochastic downscaled values are sampled from the distribution creating variance and RMSE inflation.  The results are similar for DeepESD and U-Net methods, with mean values well represented by all methods with the exception of the SQR model, which  underestimates mean values by nearly $25\%$. When analyzing the two components (rain frequency R01 and intensity SDII) separately, we found that the bias in SQR results from the intensity component. More noticeable, the MSE-based method  largely over/under-estimate  frequency/intensity, resulting in unbiased mean values (since this is the metric minimized during training) but with biased components. For the extreme values P99 and Rx1day, the STO models exhibit the best results. However, DeepESD overestimates Rx1day by a significant margin (approximately 25\%) and exhibits a large spatial variability, as shown by the wider violin plot.  All deterministic methods underestimate these metrics, with the ASYM models showing the best performance among them, with small sensitivity to the training instance. 

\begin{figure*}
    \centering
    \noindent\includegraphics[width=\linewidth]{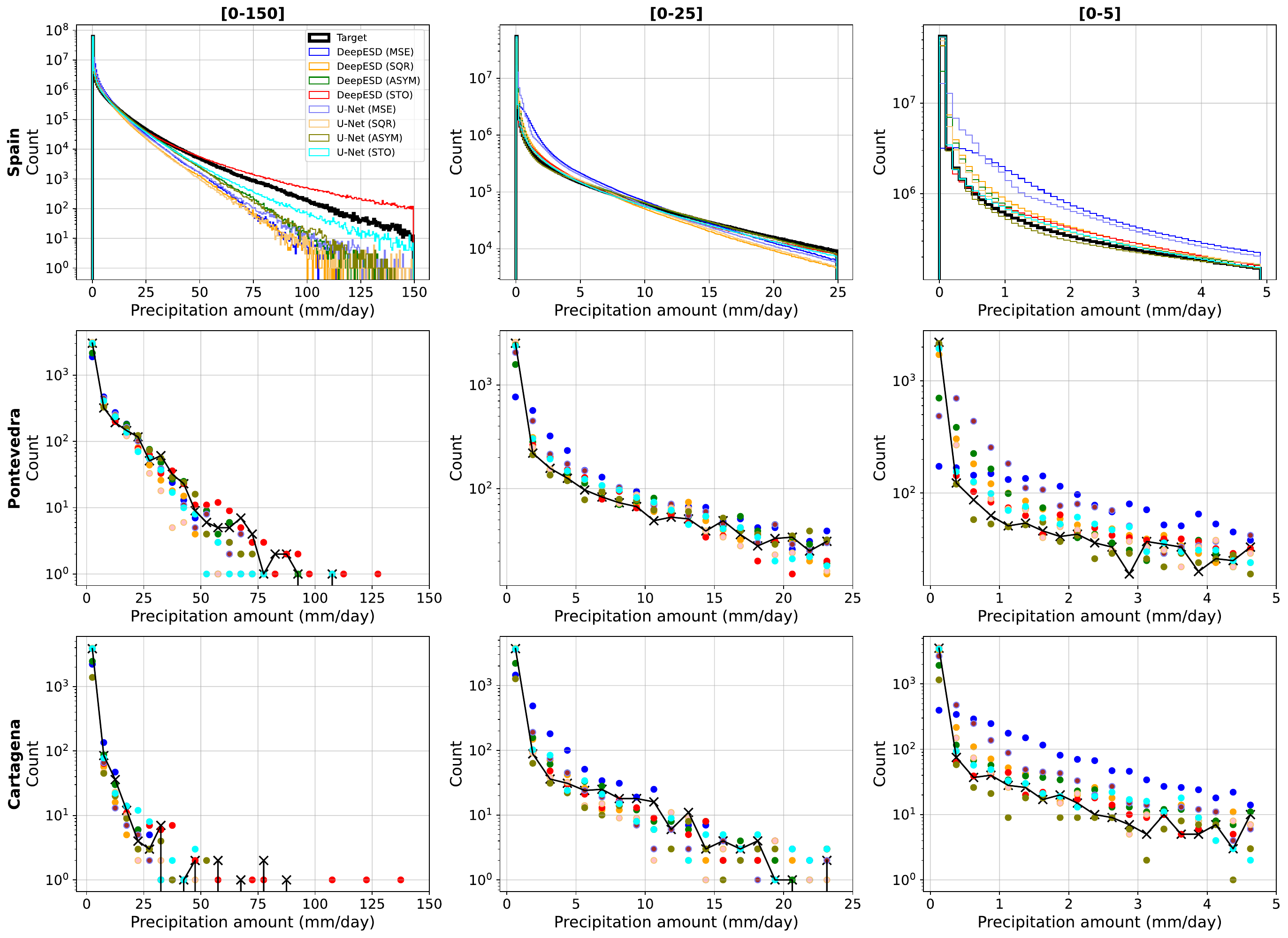}
    \caption{Histograms of precipitation for all grid points in the predictand domain (first row) and for two specific grid points, Pontevedra and Cartagena, in the second and third rows, respectively. Histograms are computed for the test set (2011-2020) of the target dataset (in black) and all the DL models intercompared (in different colors). Each column displays the histogram corresponding to the precipitation intervals 0-150 mm, 0-25 mm, and 0-5 mm. For better visualization, histogram bins for the two specific grid points are plotted as points (for the DL models) and as black crosses connected by a line for the target dataset.}
    \label{fig:fig_hist_pr}
\end{figure*}

To further analyze the representation of the precipitation distribution,  Figure \ref{fig:fig_hist_pr} displays the histograms of precipitation over the test set for all grid points in the predictand (first row) and for two specific grid points: Pontevedra and Cartagena (second and third row, respectively). For each location, histograms of the target dataset are represented in black, while histograms of the various combinations of DL architectures and loss functions are shown in different colors. The y-axis is logarithmically scaled to facilitate comparison among models. To ease visualization, we present these histograms across three different intervals: 0-150 mm, 0-25 mm, and 0-5 mm (in columns).

Focusing on the first row of Figure \ref{fig:fig_hist_pr}, which corresponds to the histogram pooling all grid points, we observe that for the most extreme values in the interval 0-150 mm, the STO-based models align most closely with the target data. Among these, DeepESD notably overestimates these extreme values by a wide margin, as previously illustrated for the Rx1day in Figure \ref{fig:fig_violin_pr}. Among the non-STO models, the ASYM-based models exhibit the best performance, especially within the 0-150 mm interval. This trend can also be observed in a less extreme interval, such as 0-25 mm (second column). Here, the STO- and ASYM-based models best follow the target data, while the other models tend to overestimate the 0-10 mm range and underestimate the 10-25 range. In the 0-5 interval, it can be observed how the MSE-based models significantly overestimate this part of the distribution while underestimating dry days. Fitting the model to the square root of precipitation (SQR) mitigates this issue, especially regarding the dry days, but still leads to the overestimation in the 0-5 mm interval. Moreover, the SQR models tend to underestimate values above 10 mm (see 0-25 interval). In contrast, ASYM and STO models accurately reproduce this part of the distribution, as well as the dry days, with STO models performing particularly well.

The second and third rows of Figure \ref{fig:fig_hist_pr} display histograms for two specific illustrative grid points: Pontevedra and Cartagena. Pontevedra, located in northwestern Spain, experiences frequent precipitation due to Atlantic humidity brought by western winds. In this case, all models behave similarly to previous observations, with DeepESD STO not overestimating extremes as much in the 0-150 interval. Additionally, DeepESD ASYM slightly underestimates the zeroes, while U-Net ASYM achieves the best results. Conversely, Cartagena, in eastern Spain, has Mediterranean conditions, with less continuous precipitation but more extreme events. Here, DeepESD STO overestimates these extremes, demonstrating that DL models adapt to different precipitation dynamics across spatial locations. For the rest of the distribution, all DL models perform as previously noted, with an improved capture of zeroes overall. Overall, ASYM downscaling methods are able to reproduce the distribution of precipitation, with the exception of very large extremes. 

\begin{figure*}
    \centering
    \noindent\includegraphics[width=\linewidth]{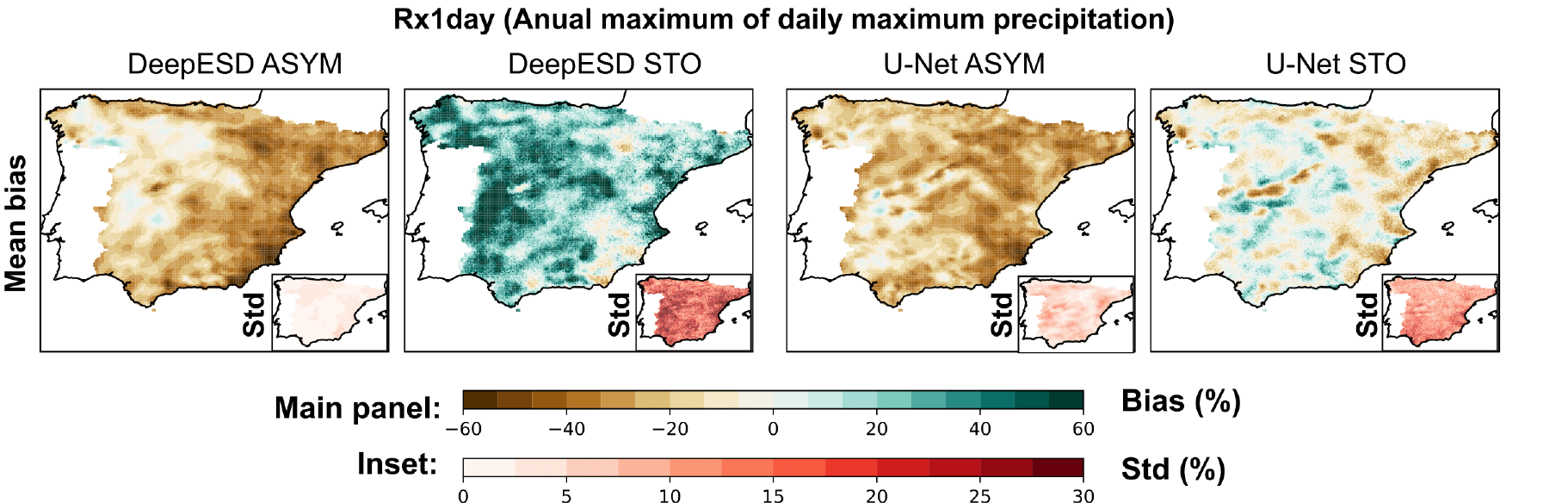}
    \caption{As Figure \protect\ref{fig:fig_extremes_tas} but for relative biases (\%) of the annual maximum of daily precipitation (Rx1day) of the STO- and ASYM-based DeepESD and U-Net models.}
    \label{fig:fig_extremes_pr}
\end{figure*}

Figure \ref{fig:fig_extremes_pr} presents the spatial results  for Rx1day precipitation metric for the DeepESD and U-Net models trained with the ASYM and STO loss functions. The ASYM versions exhibit similar patterns for DeepESD and U-Net, underestimating precipitation particularly on the eastern region. The stochastic version of DeepESD overestimate the extremes, whereas exhibits the lowest biases for the case of U-Net. However, both STO versions exhibit a large variability of results for different training replicas and, therefore, are less robust than the deterministic versions.  This variability may stem from the stochastic nature of STO-based models, which could be further exacerbated by sensitivity to the initialization of their parameters. 

\subsection{Extrapolation to Future Global Climate Projections}

From the evaluation analysis presented in the previous section, we select good performing methods to assess and intercompare the performance for downscaling  global climate projections. In particular, two methods (DeepESD MSE and U-Net MSE) have been selected for minimum/maximum temperature downscaling, and four methods (DeepESD ASYM, DeepESD STO, U-Net ASYM, and U-Net STO) for precipitation. In this section, we assess the extrapolation capabilities of these methods for future conditions, which was the main objective of this study.

We applied the methods trained and validated in the previous section to downscale the GCMs displayed in Table \ref{tab:GCMlist} for the historical and SSP3-7.0 scenarios. We selected two GCMs for temperature and two for precipitation with contrasting future climate change signals and spatial patterns, which will allow us to explore the plausibility of the downscaling methods using the raw GCM projections as pseudo-reality \citep{bano_downscaling_2022,vrac_general_2007}. For instance, for minimum/maximum temperature, the EC-Earth3-Veg model simulates a warmer future compared to the MPI-ESM1-2-LR model, whereas for precipitation, the spatial projected patterns differ between the EC-Earth3-Veg and the CMCC-CM2-SR5 models. DL-based projections are computed for a historical (1980-2014) and three different future periods (2015-2040, 2041-2070 and 2071-2100). 

\begin{figure*}
    \centering
    \noindent\includegraphics[width=\linewidth]{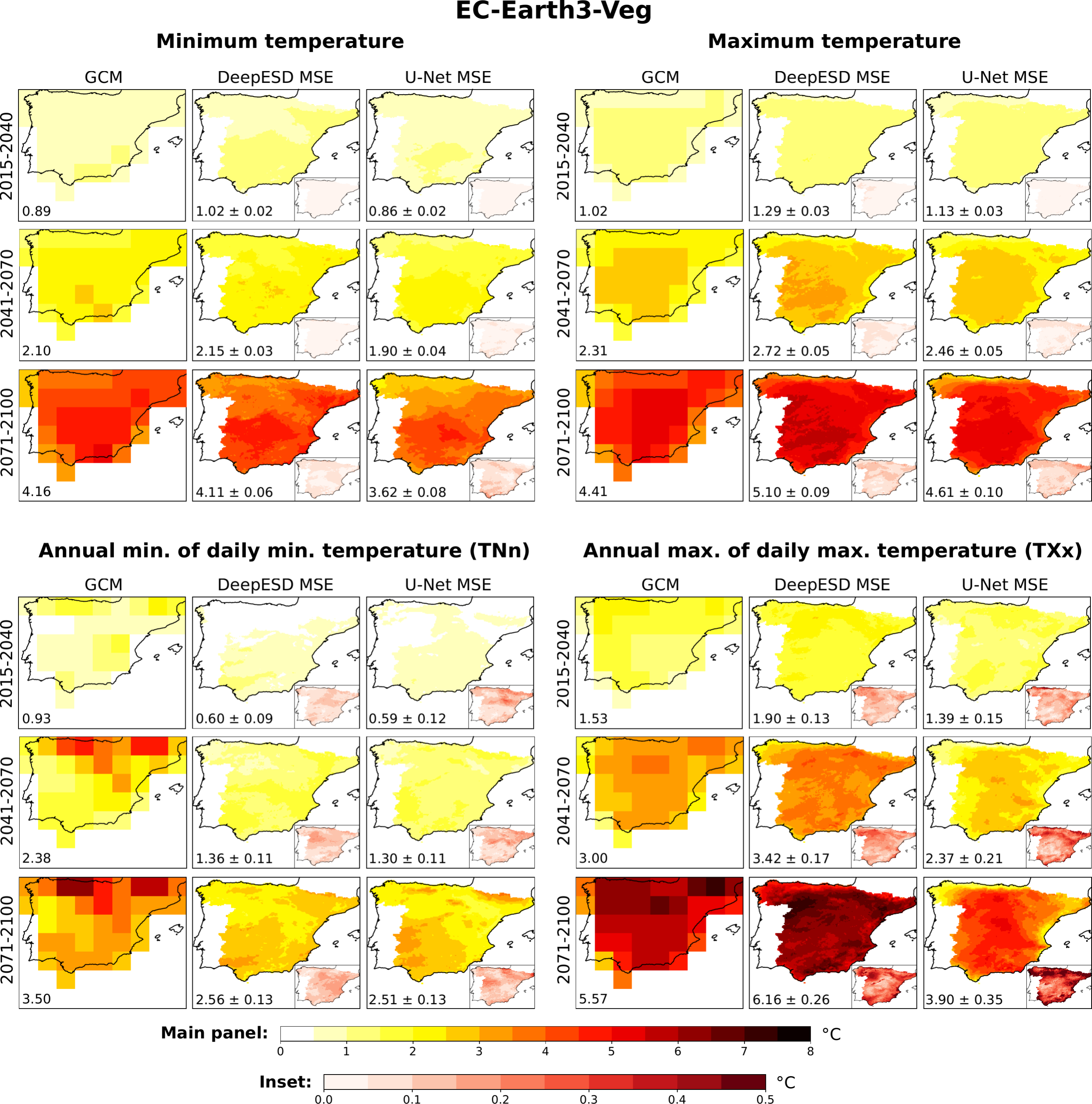}
    \caption{Climate change signals for the mean minimum and maximum temperatures (top) and the TNn and TXx indices (bottom) corresponding to the EC-Earth3-Veg climate model. For each of these, we show the signals from the GCM model and the MSE-based DeepESD and U-Net models (in columns) for three different future periods (in rows). The climate change signal shown for the DL models corresponds to the median replica. Within each subplot, the spatial standard deviation across the seven replicas is shown (inset), along with the spatial mean of the signal and the standard deviation (bottom-left).}
    \label{fig:fig_ccs_ec-earth_tas}
\end{figure*}

Figure \ref{fig:fig_ccs_ec-earth_tas} shows the climate change signal of the mean minimum and maximum temperatures and the TNn and TXx indices for the EC-Earth3-Veg climate model. For each of these indices, we show the signal from the GCM being downscaled and from the selected DL models (MSE-based DeepESD and U-Net), in columns, across three different future periods (2015-2040, 2041-2070, and 2071-2100), in rows. As introduced in Section \ref{sec:framework}, the climate change signal refers to the difference in a specific index (e.g., Mean) of the variable under analysis (e.g., daily minimum temperature) between the future and historical projections produced by the DL models. These projections are obtained by feeding the models with predictor variables from the GCM. For the DL models, the signal shown is computed from the model corresponding to the median of the seven replicas (considering the spatial mean). To assess the variability across these replicas, we show the maps with standard deviation of each signal (bottom-right within each DL subplot). We also show the numbers corresponding to the spatial mean of the signal and the standard deviation in the bottom-left of each subplot. 

For the minimum temperature, the DeepESD model aligns with the evolution of the climate change signal of the GCM along time, both in terms of the spatial pattern and the magnitude of the change (which can be assessed by focusing on the spatial mean). The U-Net produces a similar spatial pattern but, particularly in the later periods, with slightly smaller change magnitude  (up to $0.5^\circ$C), especially in the northwestern region. For the maximum temperature, the DeepESD model exhibits slightly larger  magnitude of change along time (up to $0.5^\circ$C), whereas the U-Net model aligns well with the GCM signal. Regarding the extremes, although both DL models project a similar spatial pattern for TNn, it greatly differs from that of the GCM. Both DL models translate the coarse warming signals in the northeastern region to a more localized warming over mountain areas (in particular the Pyrenees), with small intensity. In addition to these spatial differences, both DL models diverge from the GCM change trend up to $1^\circ$C. For TXx, the U-Net model does not reproduce the trend of the GCM signal, underestimating it as we move forward in time up to $1.7^\circ$C. on the other hand, the DeepESD model is closer to the GCM pattern bu slightly overestimates it. Regarding the sensitivity to the training instance, it is significantly larger for the extreme indices than for the mean, especially for the TXx. Within the TXx, the U-Net model shows higher variability among replicas, particularly in coastal regions.

\begin{figure*}
    \centering
    \noindent\includegraphics[width=\linewidth]{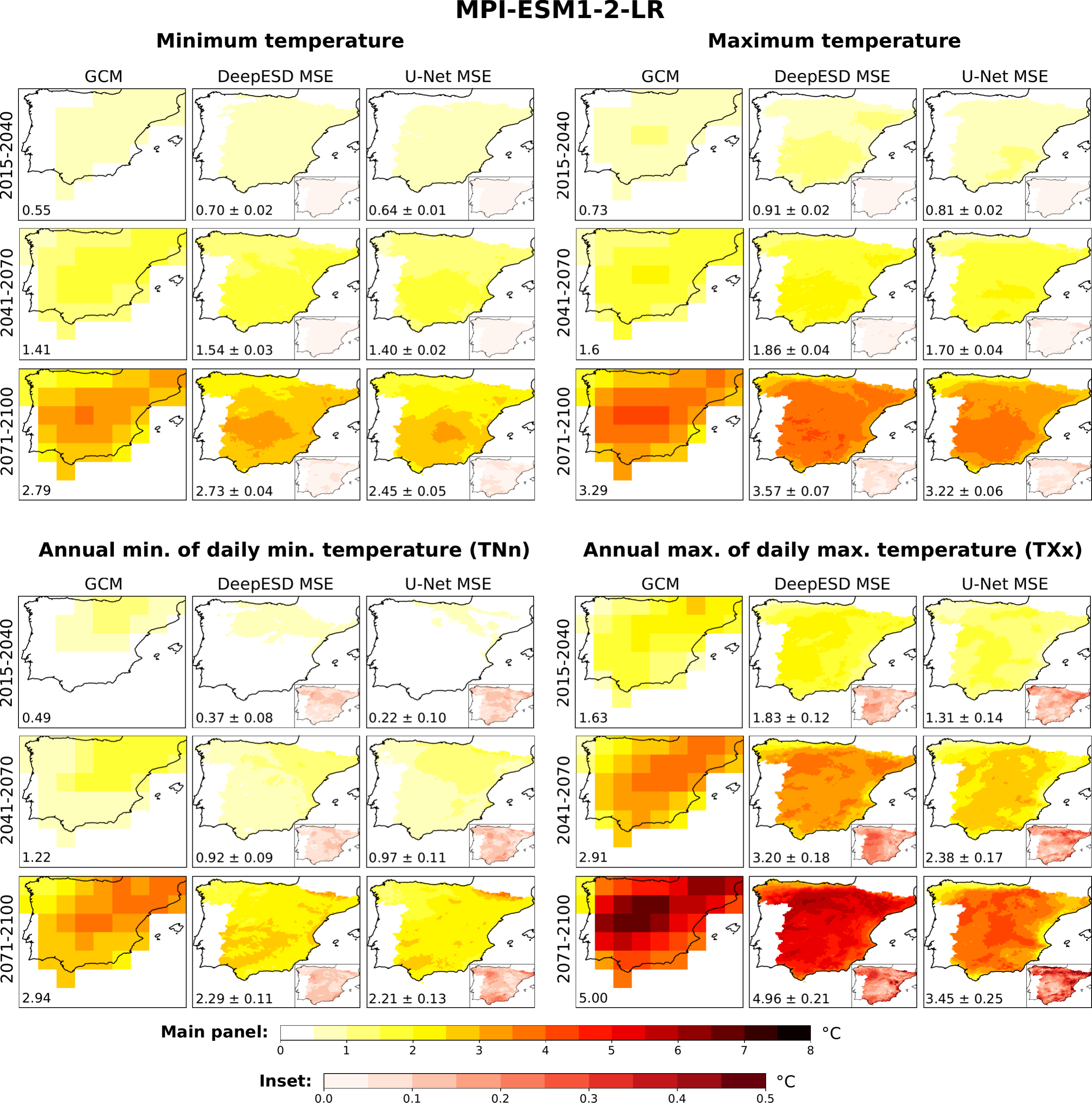}
    \caption{Same as Figure \ref{fig:fig_ccs_ec-earth_tas} but for the MPI-ESM1-2-LR climate model.}
    \label{fig:fig_ccs_mpi_tas}
\end{figure*}

In Figure \ref{fig:fig_ccs_mpi_tas}, we present the same analysis but for the downscaling of the MPI-ESM1-2-LR model, with a lower warming signal  than the EC-Earth3-Veg. The overall results are similar to the previous case, with DeepESD projecting warmer signals than the U-Net, particularly for maximum temperature. Similar to the previous GCM, the DL models do not replicate the spatial pattern for TNn, although they do project the GCM's warming of the northeastern region over the Pyrenees. Regarding the standard deviation of the different replicas, it is still larger for the extremes (in comparison to the mean), although, in this case, there is not much difference between the DeepESD and U-Net models.

\begin{figure*}
    \centering
    \noindent\includegraphics[width=0.9\linewidth]{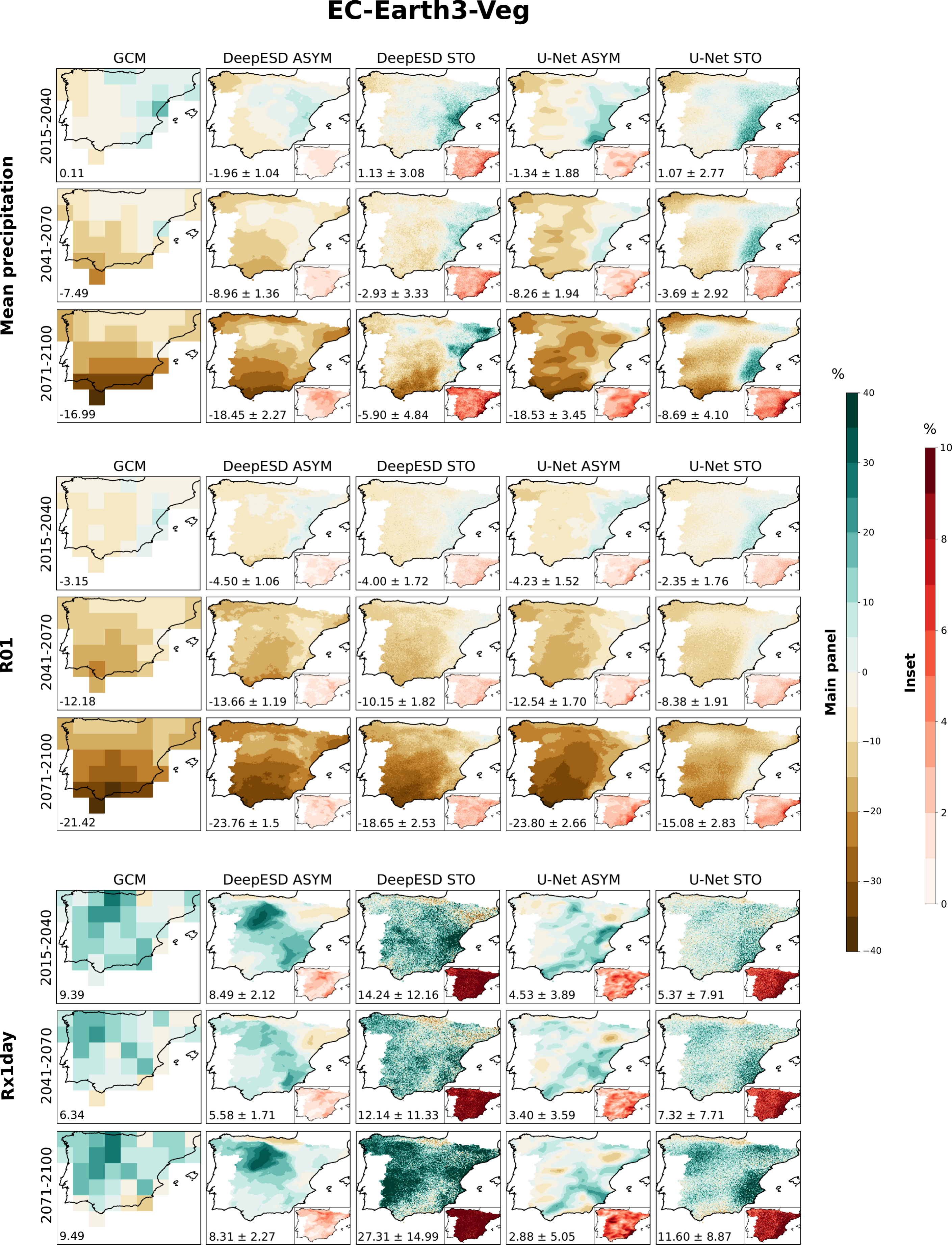}
    \caption{Same as Figure \ref{fig:fig_ccs_ec-earth_tas} but for precipitation. In this case, the climate change signal is computed for the mean precipitation, R01, and Rx1day indices (in rows). The DL models compared are the ASYM- and STO-based DeepESD and U-Net.}
    \label{fig:fig_ccs_ec-earth_pr}
\end{figure*}

Similarly to temperature, Figure \ref{fig:fig_ccs_ec-earth_pr} shows the climate change signal for mean precipitation, R01 and Rx1day indices for the EC-Earth3-Veg climate model. In this case, we compare the ASYM- and STO-based DeepESD and U-Net models. For mean precipitation in the first period (2015-2040), all DL models produce a spatial pattern similar to that of the GCM, with a general decrease, except for an increase along the eastern coast. However, as we progress in time, the STO-based models show a persistent and increasing precipitation trend along the eastern coast, whereas the GCM indicates a drying trend across the whole area of Spain. On the other hand, the ASYM-based models follow this drying trend and exhibit a similar change pattern over time to the GCM. For the R01 index, all models successfully reproduce both the spatial structure and the magnitude trends of the GCM, which simulates a reduction in the proportion of wet days over time. Regarding extremes, the Rx1day index shows more differences across DL models. The STO-based models produce a different spatial structure than the GCM, with DeepESD significantly overestimating the climate change signal. In addition, these models exhibit spatially inconsistent projections, as evidenced by the noisy structure of the climate change signal. Similarly, the U-Net ASYM model produces a spatial pattern that differs significantly from the GCM. However, the DeepESD ASYM model simulates a comparable spatial pattern over time, as seen in the increase in the north-central region of Spain during the first and last periods. Additionally, this model shows a trend more similar to that of the GCM. Regarding variability across different training replicas, the DeepESD ASYM model produces the most robust climate change signals, particularly for the Rx1day index, where STO models suffer from large variability. The index with the least variability among all DL models is the R01, reflecting the consensus among the DL models. For mean precipitation, variability is higher, especially along the eastern coast, where DL models (except for the DeepESD ASYM) deviate from the GCM.

\begin{figure*}
    \centering
    \noindent\includegraphics[width=0.9\linewidth]{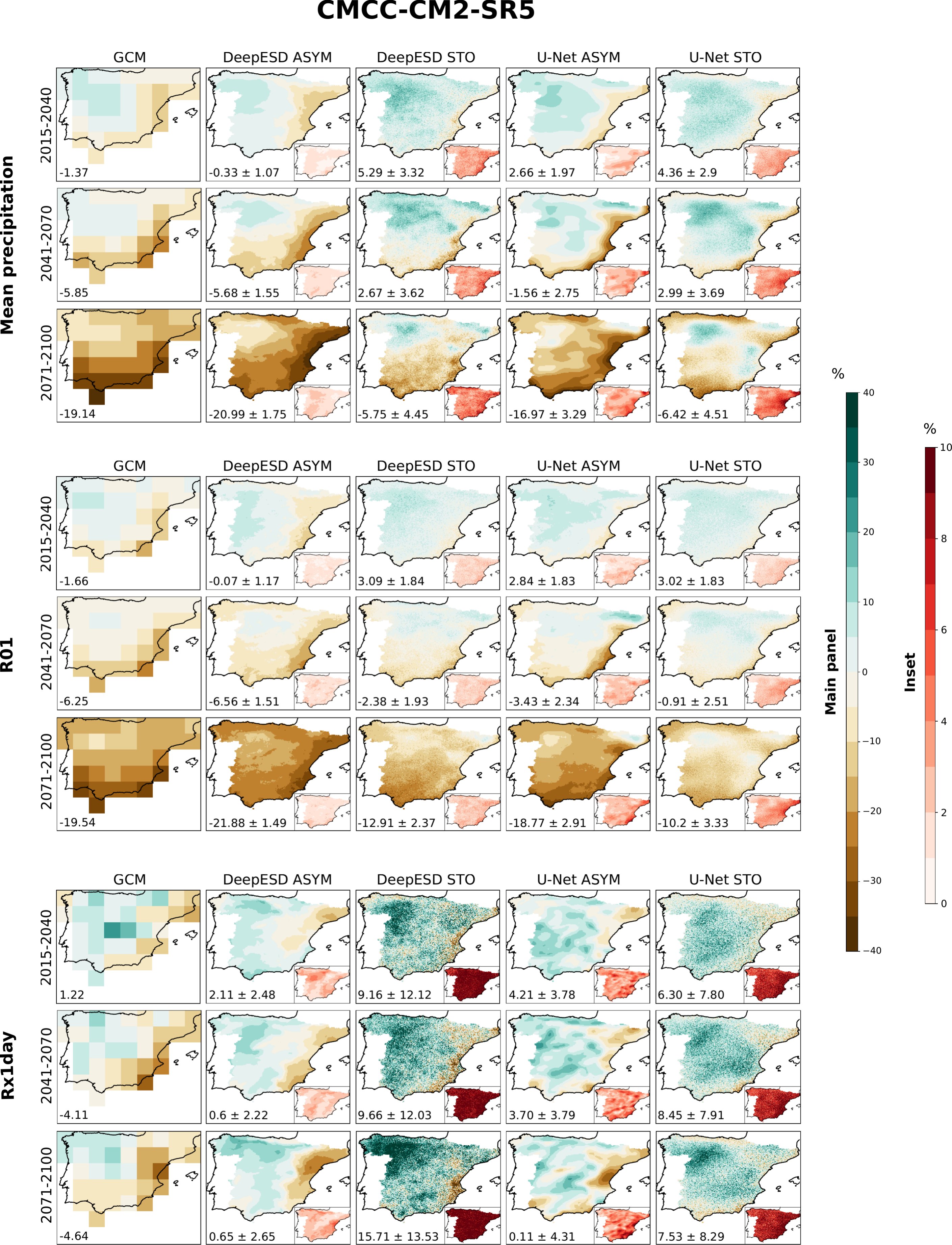}
    \caption{Same as Figure \ref{fig:fig_ccs_ec-earth_pr} but for the CMCC-CM2-SR5 climate model.}
    \label{fig:fig_ccs_cmcc_pr}
\end{figure*}

Figure \ref{fig:fig_ccs_cmcc_pr} depicts the same analysis but for the downscaling of the CMCC-CM2-SR5 climate model with similar overall conclusions. 

\section{Discussion}
\label{sec:discussion}

The results presented in this study enable a comprehensive assessment of the most popular DL downscaling models (DeepESD and U-Net), particularly regarding their ability to extrapolate to future scenarios, the primary objective of downscaling. Following the standard PP procedure, we first evaluate these models in the observational space before transitioning to future climate projections. A key part of this evaluation involves comparing the climate change signal produced by the DL models to that of the driving GCM. This is a widely used method for assessing the plausibility of DL-based projections, especially in terms of capturing large-scale trends and magnitudes \citep{bano_suitability_2021,bano_downscaling_2022,soares_high_2023}, as done in this study. While this type of comparison has its limitations—since GCMs do not reproduce the regional detail introduced by DL models—some differences in the resulting signals are to be expected and may be valid. Nevertheless, consistency in large-scale features provides a meaningful benchmark for evaluating the plausibility of the DL-derived climate change signals.

Contrary to \citet{quesada_repeatable_2022}, we do not find that U-Net models outperform DeepESD. In fact, for some specific temperature indices, DeepESD shows better performance. This discrepancy might be due to the different spatial domains of study or to the inclusion of a final dense layer in the U-Net developed in \citet{quesada_repeatable_2022}, which makes the model not fully convolutional, unlike in this study. For the different loss functions compared, we find that for temperature, both MSE and STO loss functions yield similar results, with STO performing slightly better for extremes, as previously noted in \citet{bano_configuration_2020}. However, this difference is minor, leading us to prefer the MSE-based model for its simplicity, as STO nearly doubles the model's parameters and introduce spatial heterogeneity in the sampling process. For precipitation, as noted in the literature \citep{adewoyin_tru_2021}, MSE-based models struggle to reproduce several aspects of the target distribution. Pre-processing precipitation to avoid long tails (SQR) has shown some improvements over plain MSE but still fails to properly capture the tail of the distribution. Recent approaches in the literature involve weighting extreme values in the loss function to encourage the model to better reproduce extreme values \citep{price_increasing_2022, doury_suitability_2024}. In this study, one such loss function (ASYM) provides satisfactory results across several aspects of the distribution, especially among deterministic loss functions. Another successful loss function for precipitation is STO, as previously noted in \citet{bano_configuration_2020} and observed in the present work.

In the extrapolation regime, for mean minimum and maximum temperatures, both architectures project a plausible climate change signal, broadly similar in magnitude to that of the GCM. Additionally, the spatial structure of the DL projections resembles that of the corresponding GCMs, demonstrating that the DL models can adapt to the large-scale dynamics of different climate models, an important assumption in the PP approach \citep{maraun_statistical_2018}. For extremes, specifically TNn, DL models diverge from the GCM in the spatial structure of the changes. Assessing the plausibility of these results is challenging, as they may be influenced by the local information integrated into the DL model. This raises questions about the ability of DL models to simulate changes in extremes under climate change conditions, especially in light of the TXx results. Regarding this index, for the less warm climate model (MPI-ESM1-2-LR), DeepESD seems to simulate a plausible change, though, as with TNn, the spatial structure differs. However, the U-Net model significantly underestimates the climate change signal for warmer extremes, a failing observed in both climate models. This issue with the U-Net model's ability to extrapolate to warmer extremes under climate change conditions has been noted recently in previous works in the context of emulation \citep{doury_regional_2023,hernanz_limitations_2024}. No previous works address the reason behind this behavior, but we hypothesize that it could result from the fully convolutional structure, which may constrain the potential set of functions the model can learn, leading to simpler functions unable to extrapolate to future conditions. On the other hand, for the TXx of the warmer model (EC-Earth3-Veg), the DeepESD model appears to slightly overestimate the climate change signal, a behavior previously observed for South America \citep{balmaceda_regional_2024}. This behavior could be related to extrapolation issues for warmer extremes associated with the use of dense layers, as explored in \citet{gonzalez_using_2023}. This uncertainty regarding warmer extremes is also evident in the increased variability among training replicas for TXx, where the DL models need to extrapolate the most. This suggests that some DL architectures may struggle to generalize to extreme conditions not encountered during training.

Regarding precipitation, similar to temperature, both architectures are able to capture the spatial signals of the different GCMs, reproducing the corresponding magnitudes and spatial patterns. For mean precipitation and R01 indices, all architectures and loss functions appear to simulate a plausible climate change signal. However, for the Rx1day index, some discrepancies are observed, particularly an overestimation by the DeepESD STO model. This behavior mirrors that of the same architecture for temperature (TXx). One could argue that the overparameterization caused by the dense layers might be responsible, but DeepESD ASYM, the same architecture with another loss function, does not exhibit this issue. In fact, this model performs best for simulating precipitation under climate change conditions, as it exhibits magnitudes and spatial structures closest to those of the GCM. Other possible explanations might include the difficulty of fitting appropriate probability distributions at each grid point for such high-resolution data, as previous applications of the DeepESD STO model at lower resolutions (e.g., $0.5^\circ$) did not encounter these problems \citep{bano_suitability_2021,bano_downscaling_2022,soares_high_2023}. In contrast to the excellent results of the DeepESD ASYM, the U-Net ASYM model fails to downscale precipitation, especially for the Rx1day index. This issue may be related to the underestimation of warm temperature extremes, as the fully convolutional structure might lead to overly simplistic learnt representations. For precipitation, where complex relationships between large and local scales may exist, this simplicity might be insufficient for properly modeling these relationships, particularly when extrapolating to future scenarios. Thus, for this variable, including some form of dense layers may be crucial for accurately capturing the physical phenomena, which could explain the discrepancy between the results of this work and those of \citet{quesada_repeatable_2022}.

Another key limitation of STO-based models for precipitation downscaling is that they learn independent probability distributions at each grid point, resulting in spatially inconsistent patterns when sampling. This issue is particularly evident in the noisy spatial structure of the Rx1day index, though it is also present in the Mean and R01 indices. The stronger inconsistency in Rx1day arises from its reliance on a few extreme days (annual maxima), while the Mean and R01 are computed over all days in the period, helping to smooth out spatial irregularities. This limitation of the STO loss function has been previously discussed in the literature \citep{gonzalez_use_2023,gonzalez_likelihood_2024}.

\section{Conclusions and future directions}
\label{sec:conclusions}

DL methods have shown promising results for statistical downscaling in different applications. Despite its recent emergence, numerous studies have explored this area, comparing different architectures for various variables across diverse regions. This work, for the first time, performs an exhaustive literature review to provide a global overview of state-of-the-art deep PP downscaling methods and an intercomparison based on a common experimental protocol. 

Our findings indicate that state-of-the-art downscaling methods are based mainly in fully convolutional (U-Net) or convolutional and dense models (DeepESD). When the architecture and loss functions (including value and distributional error variants)  are appropriately selected, both models can reproduce temperatures and precipitation in an evaluation period, with some systematic biases in the extremes. For precipitation, an asymmetric loss function weighting the tail of the distribution produced the best results, whereas standard MSE even over transformed (squared root) values were not suitable to reproduce the distribution of precipitation. Distributional loss functions were best in reproducing the tails of the distributions, but they exhibited the larger sensitivity to different training instances and, therefore, introduce larger uncertainty in the process. 

When applied to global projections of future scenarios, both models effectively capture the spatial patterns of temperature and precipitation changes across different climate models. DeepESD performs slightly better overall, especially for precipitation. For changes in  mean temperatures, there is a tendency of DeepESD/U-Net models to over-/under-estimate the intensity of the warming signal up to 10\% (with maximum deviations around 0.5ºC), with the exception of the distributional loss functions (which were tested only for precipitation) which do not represent the spatial pattern and fail to represent the magnitudes. 
We argue that the flexibility of the final dense layer allows DeepESD to better accommodate local downscaling in a way which can effectively extrapolate. This comes at the cost of an extra number of parameters (dense layers are densely connected), making the model more complex and less scalable for continental-wide applications. 

In the case of extreme values, for annual maximum temperature U-Net systematically underestimate warming changes up to 1.5ºC, whereas the DeepESD models follows the same behavior as with the mean values. This behavior for U-Net was previously reported in the literature in other problems \citep{doury_regional_2023,hernanz_limitations_2024}. In the case of annual minimum temperature, both methods consistently underestimate the intensity of the signal and modify the spatial pattern of the signal making it more aligned with orographic details; whether this is an added value of these methods or a lack of extrapolation remains unclear yet. 

This is an important problem within the DL field due to the lack of a theoretical framework to assess an architecture's extrapolation capability \citep{prince_understanding_2023}. A promising strategy is to pretrain these models on available GCM and Regional Climate Models (RCM) data, following recent trends in foundation models in fields such as language and computer vision \citep{bommasani_opportunities_2021}, and more recently in weather and climate modeling \citep{nguyen_climax_2023,bodnar_aurora_2024}. Such pretraining could expose DL models to a broader range of conditions, mitigating biases when transferring them to future scenarios. Additionally, it is concerning that some issues encountered during GCM downscaling (extrapolation conditions) were not detected in the standard evaluation method on the observational space. Building on previous works \citep{rampal_high_2022,gonzalez_using_2023,balmaceda_use_2024}, it is worth exploring eXplainable Artificial Intelligence (XAI) techniques to better understand the inner structure of DL models, thus easing detecting and understanding these failures. Finally, while the selected DL models produce plausible climate change signals, the deterministic nature of the loss functions means they do not quantify model uncertainty, which is crucial for informing stakeholders about confidence in future projections. Exploring uncertainty techniques for DL models, such as deep ensembles \citep{lakshminarayanan_simple_2017}, Bayesian neural networks \citep{neal_bayesian_2012}, or conformal prediction \citep{shafer_tutorial_2008}, might address this gap.

Overall, DeepESD is a suitable candidate for downscaling future projections, when trained to minimize the MSE and ASYM loss functions for temperature and precipitation, respectively. This method demonstrates potential for generating high-resolution, plausible climate change signals across different climate models, although it tends to overestimate large changes in extreme temperatures (with maximum deviations around $0.5^\circ$C compared to the global model signal). Future avenues to develop scalable deep downscaling methods could explore the flexibility of graph neural networks \citep{wu_comprehensive_2020,lam_graphcast_2022}, or transformers with attention mechanisms \citep{vaswani_attention_2017} to define scalable models while facilitating extrapolation with output-specific information. 

\clearpage
\acknowledgments
This work was partially supported by the project ATLAS (PID2019- 717 111481RB-I00) funded by MCIN/AEI/10.13039/501100011033. González-Abad acknowledges support from grant CPP2021-008510 funded by MICIU/AEI/10.13039/501100011033 and by the “European Union” and the “European Union NextGenerationEU/PRTR”, as well as from Project COMPOUND (TED2021-131334A-I00) funded by MCIU/AEI/10.13039/501100011033 and by the European Union NextGenerationEU/PRTR.

%
%
\datastatement
All the data necessary to reproduce the experiments in this study are publicly available. ERA5 data can be downloaded from the C3S Climate Data Store after registration and acceptance of the licensing terms (\url{https://doi.org/10.24381/cds.50314f4c}). Data from EC-Earth3-Veg, MPI-ESM1-2-LR, and CMCC-CM2-SR5 can be accessed through the Earth System Grid Federation (ESGF) portal (\url{https://esgf-data.dkrz.de}) by selecting the CMIP6 project and the CMIP activity. The ROCIO-IBEB dataset is available on the AEMET website (\url{https://www.aemet.es/en/serviciosclimaticos/cambio_climat/datos_diarios?w=2}); however, please note that this webpage is only available in Spanish. To download minimum and maximum temperature data, locate the tar.gz files under the section \textit{2. Rejillas ROCIO\_IBEB de temperaturas diarias extremas}. For precipitation data, refer to the section \textit{1. Rejilla ROCIO\_IBEB de precipitación}. The code used to reproduce these experiments is publicly accessible at \url{https://doi.org/10.5281/zenodo.14016954}.

%






%



\bibliographystyle{ametsocV6}
\bibliography{references}

\end{document}